\documentclass[journal=apchd5,manuscript=article]{achemso}

\usepackage{chemformula} 
\usepackage[T1]{fontenc} 

\author{Dmitry A. Zezyulin}
\email{dzezyulin@corp.ifmo.ru}
\affiliation{ITMO University, St. Petersburg 197101, Russia}
\author{Yaroslav V. Kartashov}
\affiliation{ICFO-Institut de Ciencies Fotoniques, The Barcelona Institute of Science and Technology, 08860 Castelldefels (Barcelona), Spain}
\alsoaffiliation{Institute of Spectroscopy, Russian Academy of Sciences, Troitsk, Moscow Region 108840, Russia}
\author{Dmitry V. Skryabin}
\affiliation{Department of Physics, University of Bath, Bath BA2 7AY, United Kingdom}
\alsoaffiliation{ITMO University, St. Petersburg 197101, Russia}
\author{Ivan A. \textcolor{black}{Shelykh}}
\affiliation{ITMO University, St. Petersburg 197101, Russia}
\alsoaffiliation{Science Institute, University of Iceland, Dunhagi 3, IS-107, Reykjavik, Iceland}

\title
  {\textcolor{black}{Spin-orbit coupled polariton condensates in a radially-periodic potential: Multiring vortices and rotating solitons}}


\begin{document}

\begin{tocentry}

\includegraphics[height=3.5cm]{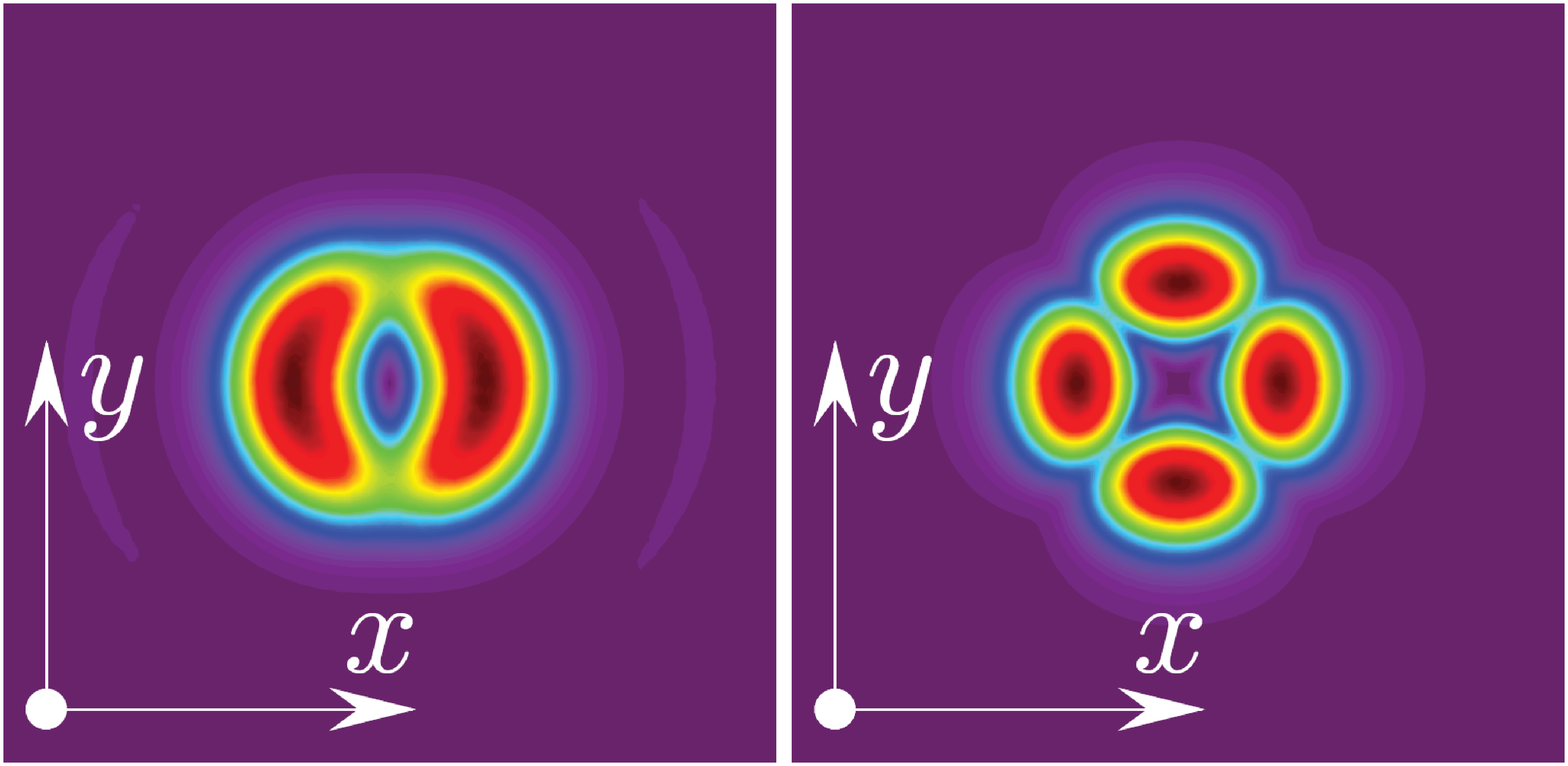}





\end{tocentry}

\begin{abstract}
 We address evolution of a spinor polariton condensate in radially periodic potentials. Such potentials allow for the observation of novel nonlinear excitations and support a variety of dynamically stable soliton states never demonstrated before in polariton condensates, including ring-like solitons with density peaks located in different radial minima of the potential and extended dynamically stable multiring patterns. Among the advantages of the system is that azimuthal modulational instabilities are suppressed due to dominating repulsive interactions between polaritons with the same spin, thereby allowing for the stabilization of radially-symmetric states. The representative feature of this system is that spin-orbit coupling between different spinor components requires them to carry different topological charges. Radially-symmetric states carrying different combinations of topological charges are discussed. Radially symmetric potentials also support stable rotating multipeaked solitons, whose properties unexpectedly depend not only on the magnitude of the rotation velocity, but also on its sign, i.e., on the rotation direction. The latter property is a consequence of spin-orbit coupling which breaks the  equivalence between  clockwise and counterclockwise rotations. The multiring structures are shown to be robust against unavoidable losses and are therefore amenable to observations with the presently available experimental techniques.  
\end{abstract}


\vspace{0.5cm}
Many-body interactions can play crucial role in quantum condensed matter systems and result in the appearance of the collective phenomena, which can not be qualitatively explained using the picture of non-interacting or weakly interacting particles. Well-known examples include such phenomena as superfluidity, superconductivity, ferromagnetism, fractional quantum Hall states, and many others. When strong inter-particle interactions are treated in the framework of the mean-field approximation, the behavior of the bosonic system can be well described by certain nonlinear wave equation for its macroscopic wavefunction \cite{Legett2001}, while the particular form of this equation is defined by the physical nature of the system. The presence of nonlinearity drastically modifies the behavior of excitations in such systems and may result in spontaneous pattern formation \cite{Schmittberger2016}, turbulence \cite{Barenghi2014}, appearance of self-sustained excitations, such as solitons \cite{KivsharBook} and topological vortex-carrying modes \cite{Dalfovo1999}. 

The standard platform for the investigation of  the coherent nonlinear phenomena in condensed matter is based on cold atoms. 
Such systems, however, require 
extremely low temperatures (in nanokelvin range). On the other hand, the use of the optical nonlinear systems frequently requires laser radiation with high peak powers, since optical nonlinearities in conventional transparent materials are usually very weak. However, there is a system which combines the advantages of the atomic Bose-Einstein condensates (BECs) and photonics, namely, cavity polaritons. They are composite light-matter quasiparticles formed by microcavity photons resonantly coupled to quantum well excitons. From their photonic component polaritons inherit extremely small effective mass ($\simeq 10^{-5}$ of the mass of a free electron) and robustness to decoherence, while the presence of the excitonic component enables strong polariton-polariton interactions resulting in giant $\chi^{(3)}$ optical nonlinearity~\cite{Gippius2007, Walker2015}. A broad variety of collective nonlinear phenomena were already reported in polariton systems. They include BEC and polariton lasing~\cite{Kasprzak2006,Galbiati2012, Bajoni2008, Sun2017}, self-organisation through multiple polariton-polariton scattering~\cite{Krizhanovskii2008}, quantised vortices~\cite{Lagoudakis2008,Tosi2012}, and conventional dark and bright solitons ~\cite{Amo2011,Sich2012,Chana2015}. Recent development of technologies of microcavity structuring~\cite{Schneider2017} enabled observation of one- and two-dimensional polariton solitons in periodic potentials~\cite{Kim2011,Tanese2013,Cerda2013,Ostrovskaya2013}.

One of the peculiarities of the polariton system as compared to systems based on cold atoms is related to spin structure of the polaritons. Similar to the cavity photons, polaritons have two possible spin projections on the structure growth axis corresponding to two opposite circular polarizations \cite{Shelykh2010,Sala2015}. For the polaritons with the in-plane momentum $\textbf{k}$ different from zero two circular polarizations become coupled by so-called TE-TM splitting, which can be described in terms of the in-plane effective $\textbf{k}$-dependent magnetic field, similarly to the case of 2D electrons with Rashba spin-orbit interaction \cite{Shelykh2009}. This effective spin-orbit coupling enables to use microcavity-based structures as building blocks for \textit{spinoptronic} devices \cite{Shelykh2004}. Moreover, as exchange term prevails in exciton-exciton scattering \cite{Ciuti1998}, polariton-polariton interactions are strongly spin anizotropic: polaritons with the same spin projectrions strongly repel, while polaritons with opposite spin projections can demonstrate weak attraction \cite{Wouters2007}. Delicate interplay between spin dynamics and polariton-polariton interactions leads to rich phenomenology of nonlinear polarization phenomena in microcavities, including polarization pattern formation \cite{Borgh2010}, spin switching \cite{Amo2010}, optical analogs of magnetically ordered states \cite{Ohadi2017}, half-solitons and analogs of magnetic monopoles \cite{Hivet2012}, half-vortices \cite{Rubo2007}, chimera states \cite{Gavrilov2018}, \textcolor{black}{polygon patterns \cite{poly2017}}, \textcolor{black}{nontrivial spin textures \cite{textures2015}}, and others. The impact of spin-orbit coupling, potential, and nonlinearity on evolution of polariton condensates was considered only recently for periodic \cite{KS16OL,KS16,Whitt2017} and parabolic \cite{SMS17} potentials.

However, particularly intriguing can be the interplay between nonlinear and spin-orbit coupling effects for polaritons in non-simply connected geometries, such as polariton rings, where it can result in new physical effects and qualitatively new types of evolution. In such settings the effects of, e.g., geometric phase may be of crucial importance, and they may lead to appearance of analogs of Aharonov-Bohm phase \cite{Shelykh2009}, topological spin Meissner effect \cite{Gulevich2016}, and chiral solitons \cite{Zezyulin2018} . 

In this paper we, for the first time to our knowledge, investigate nonlinear dynamics of cavity polaritons in a system of concentric polariton rings (radially periodic potential). It should be noted, that shaping of polariton polentials can be now routinely obtained by engineering of the profiles of non-resonant pump \cite{Astikopoulos2013}, selective metallization of the microcavity \cite{Lai2007}, its controllable etching \cite{Sala2015}, and other technologies reviewed in \cite{Schneider2017}. We show that spinor polariton condensates loaded in a radially periodic potential can form a rich variety of dynamically stable localized states \textcolor{black}{which possess a host of interesting special features. This includes the coexistence of single- and multi-ring radially-symmetric vortices having different topological charges in different spinor components. Due to the spinor nature of the system, these solutions remain localized even if the corresponding chemical potential lies in one of the spectral bands of the radially-periodic potential. Our system also supports stable azimuthally modulated solitons, which rotate steadily with constant angular velocity. In contrast to conventional scalar azimuthons \cite{azimuthons}, first introduced in focusing medium, our rotating states are vectorial and are sustained by repulsive nonlinearity that requires radially-periodic potential for radial confinement. Hence, Bragg reflection from radial structure is crucial for the existence of such modes that form in forbidden gaps of the potential. Stability of some of the rotating states in repulsive medium is unexpected too.} For solitons of the latter type, we also demonstrate that the  spin-orbit coupling results in different properties of solitons rotating in the clockwise and counterclockwise directions, that become non-equivalent in this system --- this is  an important manifestation of the new effect stemming from the interplay between the nonlinearity and spin-orbit coupling in a new geometry. \textcolor{black}{A similar nonequivalence between the rotation directions has been recently observed\cite{Zezyulin2018} for chiral solitons propagating in a quasi-one-dimensional ring under the combined action of the spin-orbit coupling and the external magnetic field (i.e., with nonzero Zeeman splitting between the two polarizations). Here we show that the external magnetic field is not a prerequisite for the mentioned asymmetry by using truly two-dimensional model where the properties of clockwise and anticlockwise solutions are demonstrated to be different even for the zero Zeeman splitting.} The increase of the angular velocity to large  values transforms rotating solitons into radially symmetric vortices \textcolor{black}{which are different depending on whether the rotation direction is clockwise or anticlockwise}.

\section*{Theoretical model} 
\label{sec:model}

\begin{figure}
	\begin{center}
		\includegraphics[width=0.4\textwidth]{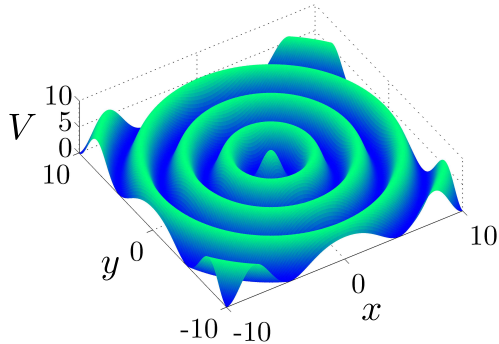}
		\caption{Radially-periodic potential $V=2V_0\cos^2(r)$, $V_0=3$, with the repelling maximum at the origin and concentric minima at $r=\pi/2, \,\, 3\pi/2,\,\, 5\pi/2, \ldots$}
		\label{fig:potential}
	\end{center}
\end{figure}

Evolution of spinor wavefunction $\boldsymbol{\psi}=(\psi_+, \psi_-)^T$ describing polaritons confined in a potential in circular polarization basis is governed by the system of 
Gross-Pitaevskii equations \cite{Flayac2010}:
\begin{equation}
\label{eq:main}
\begin{array}{l}
i \partial_t{\psi}_\pm = \left[-\frac{1}{2}\nabla^2 + V(r) + |\psi_\pm|^2 + \sigma|\psi_\mp|^2\right]\psi_\pm
+\beta(\partial_x \mp i\partial_y)^2\psi_\mp,
\end{array}
\end{equation}
where $\nabla^2 = \partial_{x}^2 + \partial_{y}^2$ is the Laplace operator, $x$ and $y$ are the spatial coordinates, $t$ is the evolution time scaled to ensure $\hbar=1$, $\beta=m^*(m_\textrm{TM}-m_\textrm{TE})/(4m_\textrm{TM}m_\textrm{TE})$ is the strength of spin-orbit coupling stemming from TE-TM splitting\cite{textures2015}, $m_\textrm{TM}$ and $m_\textrm{TE}$ are masses of polaritons polarized longitudinally (TM mode) and transversely (TE mode), and $m^* = 2m_\textrm{TM}m_\textrm{TE}/(m_\textrm{TM}+m_\textrm{TE})$ is the effective polariton mass.  Small $\sigma<0$ characterizes the attraction of spin-positive and spin-negative polaritons. Polaritons are confined in a radially-periodic potential of the form $V(r)=2V_0\cos^2(r)$, which features the local repelling maximum at the origin and multiple concentric local minima facilitating the formation of ring-shaped vortices [see Fig.~\ref{fig:potential}]. Hereafter we use polar coordinates with   radius $r$ and angle $\varphi$.

\textcolor{black}{Since the very fact of existence of solitons is not directly connected with the presence of losses and pump, here we consider a conservative system (\ref{eq:main}). The approximation of quasi-conservative dynamics of polariton condensates was used in a number of previous studies, see e.g. \cite{Sala2015,Rubo2007,Flayac2010,Gulevich2016,Zezyulin2018,textures2015,KS16,KS16OL}. In the experiment losses are expected to affect the results only quantitatively (see e.g. \cite{Walker2015}) by limitting the lifetime of the predicted states. According to recent experimental results\cite{Sun2017} the polariton lifetime can be increased to 100's of ps in high quality microcavity samples. To show that the effects predicted here are observable when dissipation is accounted for we provide below an example of propagation in the presence of linear damping.}

First we address radially symmetric solitons that have the form $\psi_\pm = \exp\{-i\mu t + il_\pm\varphi\}u_\pm(r)$, where $\mu$ is the real-valued chemical potential and real functions $u_\pm$ satisfy zero boundary conditions in the limit $r\to\infty$: $\lim_{r\to\infty}u_{\pm}=0$. Substituting these expressions in the original equations (\ref{eq:main}), one obtains that topological charges of two spinor components are not independent, but locked by the relation $l_- = l_+ + 2$. This condition appears due to the presence of spin-orbit coupling, that therefore is a crucially important physical effect determining the structure of available excitations. The functions $u_{\pm}$ satisfy the system of equations
%
\begin{equation}
\begin{array}{l}
\label{eq:fmain}
\displaystyle
\frac{1}{2}\left(\partial_r^2
+ \frac{1}{r}\partial_r - \frac{l_+^2}{r^2}\right)u_+ +(\mu - V)u_+ - (u_+^2 + \sigma u_-^2)u_+ \\[3mm]
\displaystyle
-\beta\left(\partial_r^2 + \frac{2l_++3}{r}\partial_r + \frac{l_+(l_++2)}{r^2}\right)u_-=0,\\[4mm]%
\displaystyle
\frac{1}{2}\left(\partial_r^2 
+ \frac{1}{r}\partial_r  - \frac{(l_++2)^2}{r^2} \right)u_-  + (\mu- V)u_-- (u_-^2 + \sigma u_+^2)u_- \\[3mm]
\displaystyle
-\beta\left(\partial_r^2  - \frac{2l_++1}{r}\partial_r  + \frac{l_+(l_++2)}{r^2}\right)u_+=0.
\end{array}
\end{equation}
Its localized radially-symmetric solutions can be characterized by the norms (number of quasiparticles in each component) $N_\pm = \iint |\psi_\pm|^2dxdy= 2\pi\int_0^\infty r u_\pm^2(r)dr$ and by the total norm  $N=N_+ + N_-$. Stability of stationary vortex-carrying modes can be examined using perturbed solutions in the form
\begin{equation*}
\psi_\pm = e^{-i\mu t + il_\pm\varphi}\left[u_\pm(r) + w_{\pm,q}e^{i\lambda   t + iq\varphi} + z^*_{\pm,q}e^{-i\lambda^*t-iq\varphi}\right],
\end{equation*}
where   $w_{\pm,q}$, $z_{\pm,q}$ are small radial perturbations, and integer $q =0, 1, 2, \ldots$ is the azimuthal perturbation index. Spectrum of $\lambda$ determines the growth rate of an eventual instability (the  instability corresponds to $\lambda$  with nonzero imaginary part). Linearization with respect to  $w_{\pm,q}$, $z_{\pm,q}$ yields a linear eigenvalue problem from which eigenvalues $\lambda$ can be computed numerically for each azimuthal index $q$ (in our numerics we have  examined the range $q=0,1,\ldots, 20$).

\begin{figure}
	\begin{center}
		\includegraphics[width=0.5\textwidth]{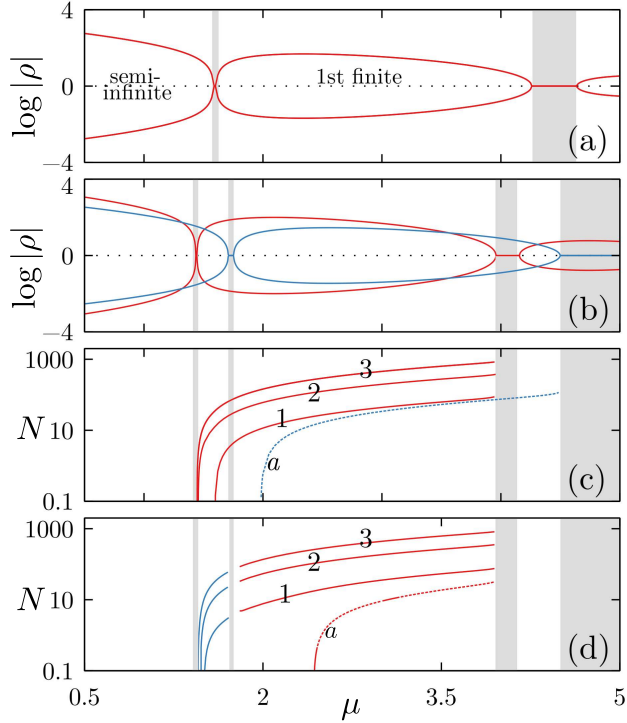}
		\caption{(a) Band-gap structure for linear periodic system with  $V_0=3$ and $\beta=0$. We show the dependencies of $\log|\rho_j|$ on $\mu$, where $\rho_j$ are the multipliers. The regions where at least one of the multipliers has unit amplitude (and the respective logarithm is zero) correspond to spectral bands (gray areas) which separate gaps (white areas). Labels ``semi-infinite'' and ``first finite'' indicate the respective gaps. (b) Band-gap structure for  $\beta=0.1$. \textcolor{black}{Blue and red curves correspond to two different pairs of multipliers which are associated with different families of solutions plotted in panels (c) and (d). } (c) Families of nonlinear modes with $l_\pm=\mp 1$. These families correspond to symmetric modes with one, two, and three ring-shaped peaks (labels 1,2,3) and antisymmetric modes with one peak  (label $a$). \textcolor{black}{Blue and black curves correspond to the blue and red curves in (b) highlighting corresponding pair of multipliers associated with a given family.} (d) Soliton families with topological charges $l_\pm =  (0, +2)$. In (c),(d) $\beta=0.1$,  $\sigma=-0.05$. For asymmetric modes solid and dashed fragments of the curves correspond to stable and unstable solutions, respectively; the stability of symmetric solutions is discussed in the text.}
		\label{fig:beta=0.1}
	\end{center}
\end{figure}

\section*{Results and discussions}

\subsection*{Linear limit} 
\label{sec:linear}

Before we proceed to the full nonlinear model, it is important to understand the properties of the system in the linear limit. The most representative property of the system with radially periodic potential is that Bragg-type reflections play substantial role in the formation of localized excitations. At $r\to\infty$, where nonlinear terms $(|u_\pm|^2+\sigma |u_\mp|^2)u_\pm$ can be neglected, and contributions from linear terms  $\sim r^{-1}, r^{-2}$ in (\ref{eq:fmain}) vanish, one obtains a truly periodic system, whose spectrum can be found using Floquet-Bloch theory if one considers $r$ as a formal variable on the entire real axis $r\in(-\infty,\infty)$. This periodic system features band-gap spectrum that can be described using the technique of the monodromy matrix \cite{YS}, i.e., the fundamental matrix of the corresponding periodic ODE system evaluated at one period (which is equal to $\pi$ in our case). Due to the spinor character of our system, the size of the monodromy matrix is $4\times 4$ . This matrix has four eigenvalues $\rho_j$ (which are called multipliers in the Floquet theory). Bloch momenta $k_j$ in the reduced Brillouin zone can be obtained from the multipliers as $k_j =-i/\pi\ln\rho_j$. Multipliers constitute two inverse pairs: $\rho_1 \rho_2 = 1$ and $\rho_3\rho_4=1$. If for some $\mu$ at least one pair of multipliers has unit absolute value ($|\rho_{1,2}|=1$ or $|\rho_{3,4}|=1$), then the linear periodic system admits a bounded solution, and this value of $\mu$ belongs to the spectral band. At the same time, the unusual property of our model is that even if $\mu$ belongs to a spectral band, the full nonlinear system can still have a soliton solution --- this situation is possible if one pair of multipliers has unitary amplitude, but another one does not (i.e., $|\rho_{1,2}|=1$ and $|\rho_{3,4}|\ne 1$ or vice versa). When all four multipliers $|\rho_j|\ne 1$, all solutions are unbounded, and this means that $\mu$ belongs to the spectral gap.

\textcolor{black}{The multipliers can be found numerically by computing directly the fundamental solution of the periodic  system over one period (with the Runge-Kutta method), constructing the $4\times 4$ monodromy matrix, and evaluating its eigenvalues.} In the absence of spin-orbit coupling (i.e., at $\beta=0$), Eq. (\ref{eq:fmain}) transforms into a pair of identical equations, so that multipliers form two identical pairs $\rho_1=\rho_3$ and $\rho_2=\rho_4$  [Fig.~\ref{fig:beta=0.1}(a)]. The spectrum of the potential at $r\to\infty$ features one semi-infinite gap and infinite number of finite gaps [white areas in Fig.~\ref{fig:beta=0.1}(a)] separated by bands [gray areas in Fig.~\ref{fig:beta=0.1}(a)]. Nonzero spin-orbit coupling $\beta$ lifts the degeneracy, and dependencies $\rho_j(\mu)$ shift with respect to each other \textcolor{black}{[see blue and red curves in Fig.~\ref{fig:beta=0.1}(b)]} that results in a more complicated band-gap structure, where several new bands emerge even at relatively small $\beta$ values [Fig.~\ref{fig:beta=0.1}(b)]. The spectrum shown in Fig.~\ref{fig:beta=0.1}(b) determines the behavior of tails of nonlinear modes at $r\to\infty$.
Since interactions between polaritons with the same spin are repulsive, one expects the formation of nonlinear modes in finite gaps. At $r\to\infty$ such modes should have oscillating tails typical for gap solitons.

The preliminary information can also be obtained for nonlinear modes of small amplitude (or, in other words, for modes with small number of particles $N$). Such \textcolor{black}{weakly  nonlinear small-amplitude} modes can be approximated by system (\ref{eq:fmain}) in   which nonlinear term $(|u_\pm|^2+\sigma |u_\mp|^2)u_\pm$ \textcolor{black}{become} negligible, but those $\sim r^{-1}, r^{-2}$ cannot be neglected anymore:
instead, they determine the asymptotical behavior of solutions close to the origin. 
Thus, one arrives at the linear eigenproblem that can be solved \textcolor{black}{numerically} at $r\in[0, \infty)$ \textcolor{black}{by discretization all derivatives and computing the eigenvalues of the resulting sparse matrix.}  The corresponding  eigenvalues $\mu$ determine the chemical potentials of localized linear modes from which the small-amplitude nonlinear modes bifurcate. The remarkable property of the potential considered here is that it supports various localized linear modes with topological charges $l_\pm = (-1, +1)$ and $l_\pm =  (0, +2)$, whose eigenvalues $\mu$ are situated \textit{inside} spectral gaps of the periodic problem discussed above. Such modes automatically acquire decaying oscillating tails in accordance with the above analysis. Usually, the further is the maximum of the linear mode from the origin, the closer its eigenvalue to the edge of the forbidden gap. The number of these modes increases with increase of the potential depth $V_0$.  
\textcolor{black}{Linear modes obtained in this way can be used as initial guesses for the iterative Newton method which converges to truly nonlinear modes described by the full system (\ref{eq:fmain}) without any neglected terms.}

\subsection*{Vortex solitons}
\label{sec:symmetric}

\subsubsection*{Solitons with $l_\pm=(-1, +1)$}

\textcolor{black}{Using the preliminary  information on the shapes of possible localized solutions obtained in the linear limit as the initial approximation, we have performed the numerical continuation of linear modes into the nonlinear domain. To this end, the standard finite-difference approach was employed  which consists in the  approximation of the derivatives with finite differences and solving the resulting system of nonlinear algebraic equations with the iterative Newton's method.}

\begin{figure}
	\begin{center}
		\includegraphics[width=0.5\textwidth]{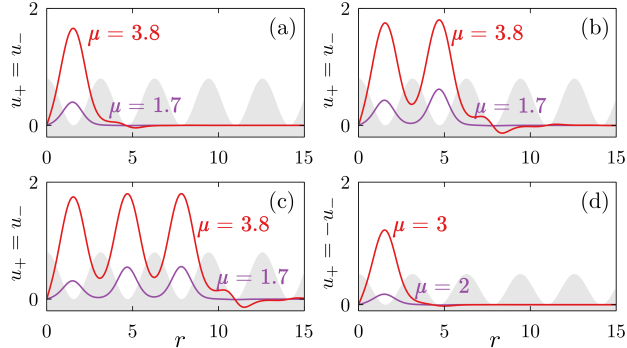}
		\caption{Representative profiles of vortex solitons with topological charges $(-1,1)$. (a,b,c) Symmetric modes with one, two, and three rings, and (d) antisymmetric mode with one ring. Corresponding $\mu$ values are shown next to the curves. Radial potential is schematically shown as shaded region.}
		\label{fig:modes_m1p1}
	\end{center}
\end{figure}

For $l_\pm = (-1, +1)$ the nonlinear system (\ref{eq:fmain}) admits two obvious reductions: $u_+=u_-$ (symmetric modes) and $u_+=-u_-$ (antisymmetric modes). Solitons with such topological charges bifurcate from corresponding symmetric or antisymmetric linear modes. The most fundamental solutions are the symmetric modes $u_+=u_-$ that feature a single bright ring residing at the first minimum of the potential (i.e., local minimum at $r=\pi/2$). Representative profiles of such solitons are shown in Fig.~\ref{fig:modes_m1p1}(a) for parameter values $V_0=3$, $\beta=0.1$, $\sigma=-0.05$ that will be used throughout this paper. Interestingly, in addition to single-ring solitons, radially-periodic potential also supports \emph{multi-ring solitons} featuring density peaks on each inner radial minimum of the potential [see Fig.~\ref{fig:modes_m1p1}(b) and (c) for examples of symmetric two-ring and three-ring solutions]. With increase of chemical potential, the amplitudes of peaks residing on different potential minima become practically equal [compare magenta and red curves in Fig.~\ref{fig:modes_m1p1}(b,c)]. As shown in Fig.~\ref{fig:beta=0.1}(c) the families of such symmetric solitons bifurcate from linear modes in the first finite gap and cease to exist at the right edge of the second gap, where such states become delocalized because the amplitude of the pair of multipliers $\rho$ associated with these solutions becomes unitary. The corresponding $N(\mu)$ dependencies in Fig.~\ref{fig:beta=0.1}(c) cross one narrow spectral band. We emphasize that inside this band the nonlinear modes \emph{remain localized}, that is possible due to the presence of two pairs of coexisting Floquet multipliers as explained above. \textcolor{black}{Since the   decaying   tails of our solutions are determined by the Floquet multipliers, each family of localized solutions plotted in Fig.~\ref{fig:beta=0.1}(c) is associated with some pair of the multipliers plotted in Fig.~\ref{fig:beta=0.1}(b). This correspondence between  Fig.~\ref{fig:beta=0.1}(b) and Fig.~\ref{fig:beta=0.1}(c) is illustrated by red and blue colors used to plot the corresponding curves.}

\begin{figure}
	\begin{center}
		\includegraphics[width=0.5\textwidth]{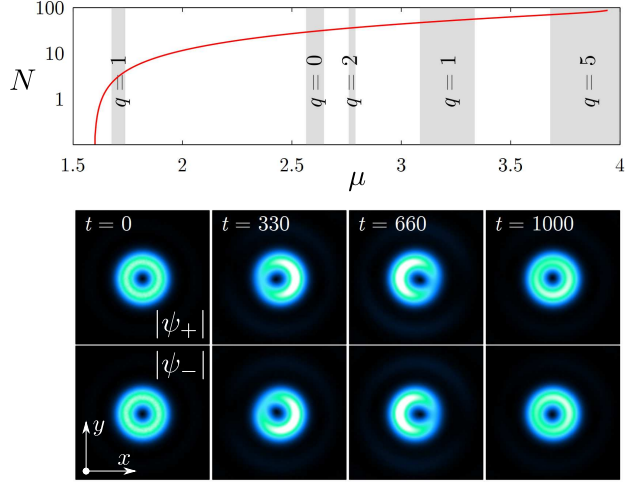}
		\caption{Upper panel shows the stability chart for symmetric vortex solitons with topological charges $(-1,1)$ and single ring-shaped peak located in the first potential minimum. Gray intervals in $\mu$ demarcate the domains of instability, while labels $q=0,1,\ldots$ indicate the azimuthal index of the most destructive perturbation. Lower panels show time evolution of the most unstable mode (which corresponds to $\mu=1.7$ and  $q=1$). Corresponding distributions are shown within $x,y\in [-6,+6]$ windows.}
		\label{fig:stab13}
	\end{center}
\end{figure}

\begin{figure}
	\begin{center}
		\includegraphics[width=0.5\textwidth]{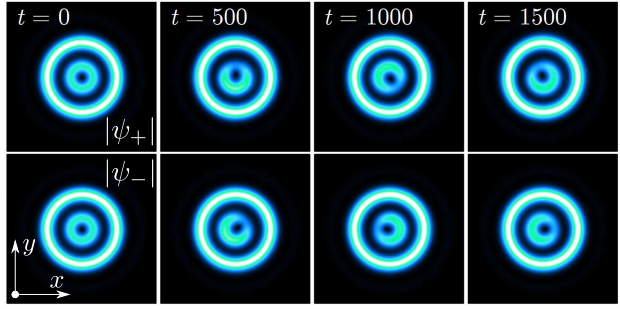}
		\caption{Time evolution of the most unstable two-peak symmetric mode with  $\mu=1.67$ and topological charge $(-1,1)$. The $x,y\in [-10,10]$ window is shown. Upper and lower rows show $|\psi_+|$ and $|\psi_-|$, respectively.}
		\label{fig:evolut10}
	\end{center}
\end{figure}

Linear stability analysis yields rather complicated structure of alternating stability and instability domains even for simplest symmetric solitons. Typical example of stability chart demarcating stable and unstable single-ring symmetric solitons is shown in the upper panel of Fig.~\ref{fig:stab13}. In all instability domains (see gray regions with indicated on them azimuthal indices of the most destructive perturbation modes) the perturbation growth rates are small ($\sim 10^{-2}$ or less). \textcolor{black}{In order to validate the predictions of the linear stability analisys, we have simulated the dynamics of  stable and unstable stationary vortex modes  using the time-dependent system (\ref{eq:main}). The latter was solved using the split-step method which efficiently approximates the Laplace operator using the fast Fourier transform. The initial conditions for system  (\ref{eq:main}) were taken in the form of stationary ring-shaped vortex modes whose profiles were perturbed with small-amplitude random noise whose maximal magnitude was about $5\%$ of the maximal amplitude of the solution.} Stable ring-like solitons persist over extremely long time intervals. Typical evolution of unstable symmetric soliton is illustrated in the lower panels of Fig.~\ref{fig:stab13}. In the course of instability development different spinor components acquire crescent-like shapes that rotate and slightly oscillate in amplitude, leading to quasi-periodic reconstruction of the ring-like structure. For two-ring and three-ring symmetric families the stability properties are similar: they can be \emph{stable} within multiple intervals of $\mu$ separated by instability domains. Evolution of the most unstable two-ring mode is shown in Fig.~\ref{fig:evolut10}. The instability mainly affects the inner ring, which transforms into rotating crescent-like distribution, while outer ring does not show any visible distortion.

In Fig.~\ref{fig:beta=0.1}(c) we also show a family of the simplest antisymmetric vortex solitons with the density peak in the first potential minimum. In contrast to symmetric solitons, this family originates in the second finite gap. Interestingly, the right border of the existence domain for antisymmetric branch also differs from that for symmetric branch. This is explained by the properties of the underlying linear periodic system (at $r\to \infty$): symmetric and antisymmetric families are associated with different pairs of multipliers (compare red and blue dependencies in Fig.~\ref{fig:beta=0.1}(b). Antisymmetric solutions [dashed curve in Fig.~\ref{fig:beta=0.1}(c)] are typically unstable, except for modes having very small amplitudes, hence we do not discuss them here in detail.

\subsubsection*{Solitons with $l_\pm=(0, +2)$}

\begin{figure}
	\begin{center}
		\includegraphics[width=0.5\textwidth]{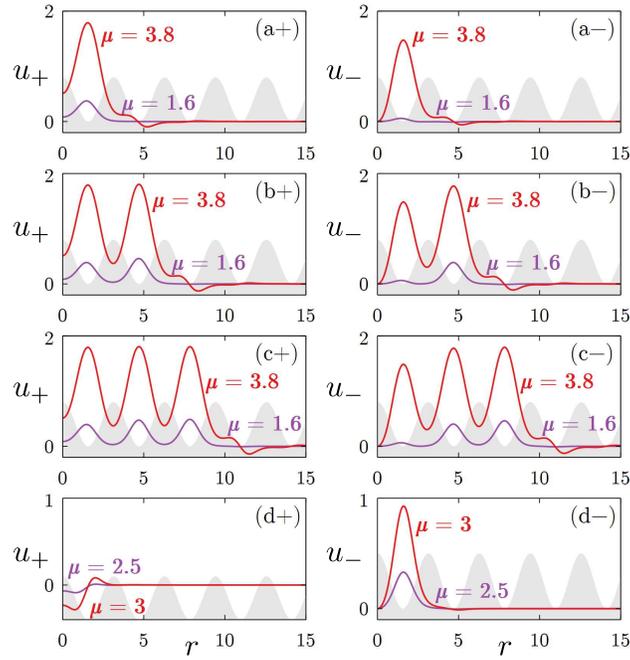}
		\caption{Representative profiles of solitons with charges $(0,+2)$. (a,b,c) Symmetric modes with one, two, and three rings, and (d) antisymmetric mode with one ring. Panels with ``$+$'' and ``$-$'' show $u_+$ and $u_-$, respectively.}
		\label{fig:modes_m0p2}
	\end{center}
\end{figure}

Next we consider vortex solitons with topological charges $l_+=0$ and $l_-=2$. In this case two spinor components have essentially different shapes. However, one can still loosely classify some of these solutions as ``symmetric'' (two components have the same sign almost everywhere) and ``antisymmetric'' (two components are typically of the opposite signs). Four families of solitons with charge $(0,+2)$ (including three ``symmetric'' families and one ``antisymmetric'' family) are shown in Fig.~\ref{fig:beta=0.1}(d), \textcolor{black}{where we again use blue and red colors to illustrate the correspondence between the nonlinear modes and the pairs of Floquet multipliers plotted in Fig.~\ref{fig:beta=0.1}(b).} The  corresponding spatial profiles of $u_+$ and $u_-$ are illustrated in Fig.~\ref{fig:modes_m0p2}. Notice different asymptotical behavior of spinor components at $r\to 0$. For symmetric solutions [Fig.~\ref{fig:modes_m0p2}(a,b,c)] the components $u_+$ and $u_-$ are typically of approximately equal amplitudes, while for the antisymmetric family the second component always dominates [$|u_-| \gg |u_+|$ in Fig.~\ref{fig:modes_m0p2}(d)]. Symmetric  $(0,+2)$ solutions can be stable (with possible bands of weak instabilities), while the antisymmetric family is chiefly unstable, but has two distinctive stability intervals: one for small-amplitude solutions and another one for solutions of moderate amplitudes [solid and dashed fragments of the corresponding curve in Fig.~\ref{fig:beta=0.1}(d)].

\subsubsection*{Effect of losses}

\begin{figure}
	\begin{center}
		\includegraphics[width=0.5\textwidth]
		{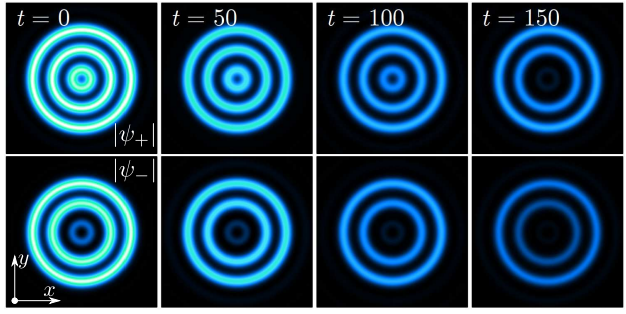}
		\caption{Decay of a multi-peak $(0, +2)$ charged vortex under the action of weak losses.  Upper and lower panels show $|\psi_+|$ and $|\psi_-|$, respectively  within $x,y\in [-12, +12]$  window.}
		\label{fig:losses}
	\end{center}
\end{figure}

The system (\ref{eq:main}) is valid only in the quasi-conservative approximation. In real polariton condensate one has to take into account the unavoidable losses. In order to address their impact on stable multi-ring vortex solitons predicted here, we have simulated their evolution in the dissipative extension of the system (\ref{eq:main}), where terms $-i\gamma \psi_+$ and  $-i\gamma \psi_-$ were added to the first and second equations, respectively. The coefficient $\gamma=0.01$ was selected in accordance with typical lifetime of polariton condensates in microcavities. The stability intervals for multi-ring solitons found here are sufficiently broad, so that gradual decrease of peak amplitude leads only to adiabatic transformation of the mode shape, usually without driving the mode from its stability domain. This is illustrated on the example of three-ring vortex solitons with charge $(0, +2)$, see Fig.~\ref{fig:losses}. This observation suggests that unavoidable dissipation does not hinder the existence of multi-ring vortices.

\subsection*{Rotating solitons} 
\label{sec:rotating}
It was shown above that unstable radially-symmetric vortex solitons may transform into quasi-stationary rotating states. The signatures of this behavior are visible in Figs.~\ref{fig:stab13} and~\ref{fig:evolut10}. This suggests that our system can support a new type of rotating solitons. In order to find them, we move into rotating coordinate frame \cite{dvs} $x'=x\cos(\omega t) + y\sin(\omega t)$, $y'=y\cos(\omega t) - x\sin(\omega t)$, where $\omega$  is the rotation frequency, and search for stationary solitons in this frame in the form $\psi_\pm = u_\pm(x', y')\exp(\mp i\omega t - i\mu t)$, where spinor components $u_\pm$ satisfy the equation: 
\begin{equation}
\label{eq:rot}
\begin{array}{c}
\mu u_\pm = -(1/2)(\partial_{x'}^2 +\partial_{y'}^2)u_\pm  + V(r)u_\pm + (|u_\pm|^2 + \sigma |u_\mp|^2)u_\pm\\[2mm]
+\beta (\partial_{x'} \mp i\partial_{y'})^2u_\mp  \mp \omega u_\pm + i\omega(x'\partial_{y'} - y'\partial_{x'})u_\pm.
\end{array}
\end{equation}
\textcolor{black}{It should be mentioned here that for $\beta \neq 0$ the transition to the rotating coordinate frame in Eq. (1) results in the appearance of additional multipliers $\sim \exp(\mp 2i\omega t)$ in front of terms describing spin-orbit coupling. Therefore, in order to eliminate time-dependence from the equation one has to assume solutions in the form $\psi_\pm = u_\pm(x', y')\exp(\mp i\omega t - i\mu t)$ that leads to the appearance of one more term $\mp \omega u_\pm$ in the Eq. (3) in addition to standard (last) Coriolis term. This new term is analogous  to the Zeeman splitting induced by the magnetic field. It should be stressed that for $\beta=0$ (in the absence of spin-orbit coupling) the substitution $\psi_\pm = u_\pm(x', y')\exp( - i\mu t)$ would be already sufficient for elimination of time-dependence in the equation in the rotating coordinate frame. Equation (\ref{eq:rot})  clearly shows that for nonzero spin-orbit coupling $\beta\ne 0$ the sign and magnitude of the angular velocity $\omega$ can affect significantly the properties of the rotating solutions.}

\begin{figure*}
	\begin{center}
		\includegraphics[width=\textwidth]{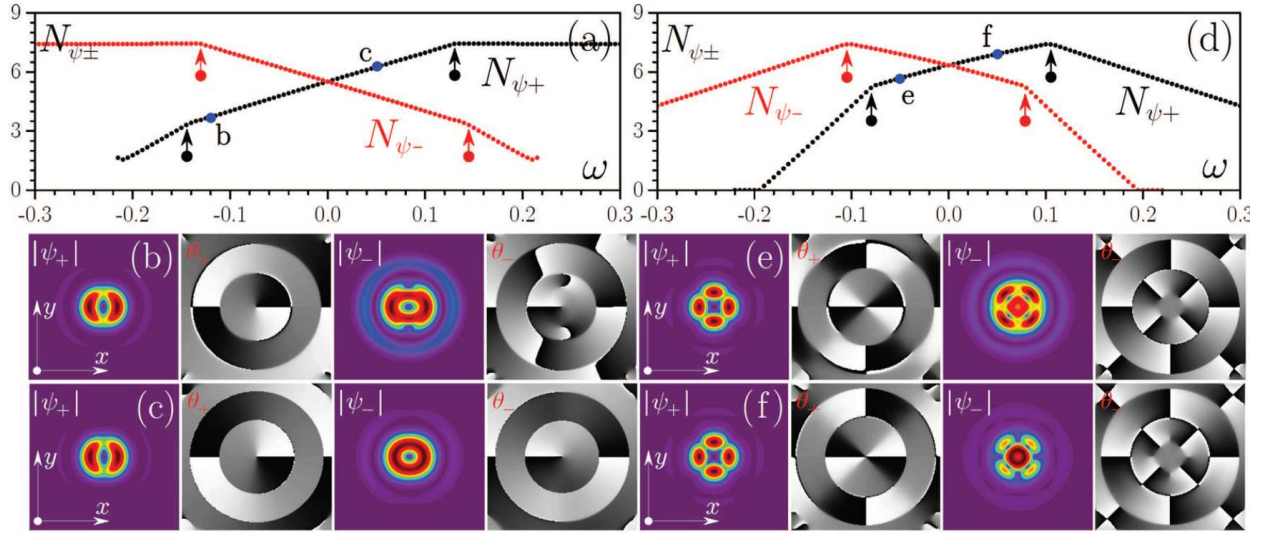}
		\caption{Norm versus rotation frequency in two-peak (a) and four-peak (d) rotating states with dominating spin-positive $(N_{\psi+})$ or spin-negative $(N_{\psi-})$ components. The interval between black (red) arrows indicates existence domains for rotating states with dominating spin-positive (spin-negative) component. Examples of two-peak states at  $\omega=-0.12$ (b), $\omega=0.05$ (c) corresponding to the points \textbf{b} and \textbf{c} in panel (a), and of four-peak states at $\omega=-0.05$ (e), $\omega=0.05$ (f) corresponding to the points \textbf{e} and \textbf{f} in panel (d). Panels (a)--(c) correspond to $\mu=2.3$, panels (d)--(f) correspond to $\mu=3.0$. Profiles are shown within $x,y\in [-8, +8]$ window.}
		\label{fig:rot1}
	\end{center}
\end{figure*}

\textcolor{black}{Equation ~(\ref{eq:rot}) supports a rich  variety of rotating solitons having different numbers of peaks and found by us numerically using a Newthon method. The simplest states feature two or four peaks that reside in the first minimum of radial potential, see Figs.~\ref{fig:rot1}(a)-(c) and Figs.~\ref{fig:rot1}(d)-(f), respectively. At $\omega= 0$ they transform into dipole and quadrupole solitons. Either  spin-positive or spin-negative component can be dominant in these states. To distinguish these two cases we denote total norms of corresponding solitons as $N_{\psi+}$ and $N_{\psi-}$, respectively. As one can see from Fig.~\ref{fig:rot1}(a) for the two-peak solitons and from Fig.~\ref{fig:rot1}(d) for the four-peak solitons, at fixed chemical potential $\mu$ these two families feature completely different dependencies $N_{\psi+}(\omega)$ (black dots) and $N_{\psi-}(\omega)$ (red dots) of norms on the rotation frequency $\omega$. One of the central results of this work is that these dependencies are \emph{asymmetric} in frequency $\omega$. This means that counterclockwise $(\omega>0)$ and clockwise ($\omega<0$) rotating solitons with the same energy  $\mu$ and same $|\omega|$  from either $N_{\psi+}$ or $N_{\psi-}$ families have different shapes and properties --- a phenomenon not reported before in any nonlinear radially symmetric setting. This asymmetry  is a result of an interplay between the rotation and spin-orbit effects.}

\textcolor{black}{Rotating solitons presented here are somewhat reminiscent of azimuthons, which have been reported earlier in scalar configurations in a self-focusing medium \cite{azimuthons} and in spinor condensates with the Josephson coupling, rather than with spin-orbit one \cite{Li2016}. In contrast to conventional azimuthons our structures are supported by repulsive nonlinearity and localized radially due to Bragg reflection from radially-periodic potential.  Two- and four-peak rotating solitons with dominating  $|\psi_+|$ component exist in the frequency interval located between the two black arrows, see Figs.~\ref{fig:rot1}(a)~and~\ref{fig:rot1}(d) [their counterparts with dominating $|\psi_-|$ component exist between the red arrows]. At the boundaries of these intervals rotating solitons with azimuthally-modulated shapes transform into 
radially-symmetric vortex solitons.}
\textcolor{black}{Thus, two-peak solitons with the dominating $|\psi_+|$ component transform into vortices with $l_\pm = (1, 3)$ at the left black arrow, while at the right black arrow these states turn into vortices with $l_\pm =(-1,1)$. Recall, that at $\omega=0$ the two-peak family turns into the dipole solutions. 
}

\textcolor{black}{Four-peak state with dominating  $|\psi_+|$ component turns into vortex with $ l_\pm = (2, 4)$ at the left black arrow and into vortex with $l_\pm=(-2, 0)$ at the right black arrow, passing through quadrupole state at $\omega=0$. Solitons with dominating  $|\psi_-|$ component show similar transformations. Notice that we deliberately did not stop soliton families at black or red arrows and continued them to larger and smaller frequencies, even though these families beyond intervals limited by the arrows correspond to radially symmetric vortex modes. These vortex modes are \textit{asymmetric}, i.e., $|\psi_+|\ne |\psi_-|$, and their norm also depends on the rotation frequency $\omega$  [with the only exception for $l_\pm  = (-1, 1)$ states where Coriolis terms cancel $\mp \omega u_\pm$ terms in (\ref{eq:rot}) and the frequency $\omega$ has no effect on the soliton].}

\begin{figure}
	\begin{center}
		\includegraphics[width=0.5\textwidth]{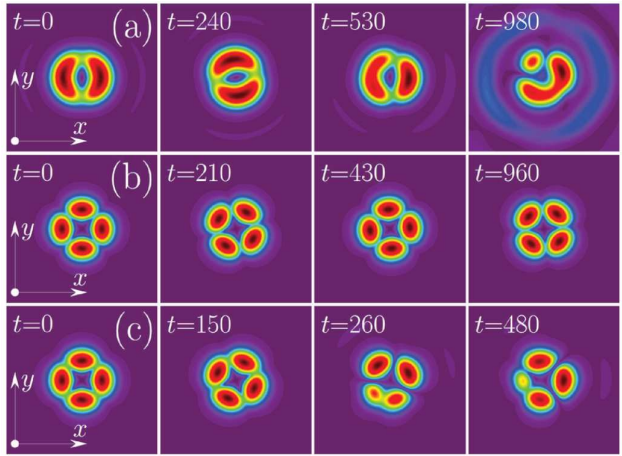}
		\caption{Evolution dynamics of the (a) unstable two-peak soliton with $\mu=2.3$, $\omega= 0.06$, (b) stable four-peak soliton with $\mu=3.0$, $\omega= 0.04$, and (c) unstable four-peak soliton with $\mu=3.0$, $\omega=0.06$. Only dominating $|\psi_-|$ component is shown within $x,y\in [-6, +6]$  window.}
		\label{fig:rot2}
	\end{center}
\end{figure}

\textcolor{black}{Stability of rotating solitons was analyzed using direct solution of Eq. (1) up to  $t\sim 10^3$ and with inputs in the form of  perturbed exact rotating states. All two-peak rotating solitons are unstable [see Fig.~\ref{fig:rot2}(a) with typical decay dynamics], but instability is suppressed close to frequency values, where such solitons turn into radially symmetric vortex modes (the latter modes are completely stable). In contrast, four-peak modes are stable in the largest part of their existence domain in $\omega$. For $\mu= 3$ such solitons with dominating spin-positive component are stable at  $-0.08 < \omega < 0.03$  (interestingly, their counterparts with  dominating $|\psi_-|$  are
stable at  $- 0.03<\omega < 0.08$). Stable and unstable evolution of four-peak states is illustrated in Figs.~\ref{fig:rot2}(b),(c). Examples of stable rotating states with larger number of peaks were found too. Stability of the rotating state with more complex internal structure is a surprising result.}

\section*{Conclusion}
\label{sec:conclusion}

To conclude, we have provided a systematic study of spin-orbit-coupled polariton states loaded in a radially-periodic trap which represents a potential in the form of a system of concentric rings. The system is demonstrated to support a rich variety of dynamically stable spinor stationary localized vortex states characterized by different topological charges in each component. Apart from the simplest solution with a single distinctive ring-shaped density peak, the system supports multi-ring states states that nevertheless can be dynamically stable despite their complex internal structure. Additionally, the radially-periodic potential enables the steady propagation of multipeaked  stable rotating solitons. Spin-orbit coupling results in the nonequivalence between the solitons rotating clockwise and counterclockwise.

Our results open several directions related to further investigation of polariton  systems trapped in circular landscape potentials. Those for instance include the analysis of the effects stemming from the additional magnetic field  which has not been taken into account herein, and can result in nontrivial topological or chiral effects \cite{Gulevich2016,Zezyulin2018}. 
From the nonlinear optics perspective, it also appears promising to study the possibility of nonlinearity-induced symmetry breaking bifurcations with the spontaneous formation of asymmetric (neither symmetric nor antisymmetric) spinor modes with topological charges $l_\pm = \mp 1$.  A more detailed investigation of the dissipative dynamics of trapped spinor vortices coupled to  reservoirs of incoherent exitons \cite{Yulin16} is also in order.

\textcolor{black}{
\section*{Author Contributions}
All co-authors contributed equally to this work.
}

\begin{acknowledgement}
D.A.Z.  was financially supported by Goverment of Russian Federation (Grant 08-08).  Y.V.K. acknowledges support from the Severo Ochoa
program (SEV-2015-0522) of the Government of Spain, Fundacio Cellex, Generalitat de Catalunya and CERCA.    I.A.S. acknowledges support from the Rannis project 163082-051, mega-grant 14.Y26.31.0015 and goszadanie 3.2614.2017/4.6 of the Ministry of Education and Science of the Russian Federation. 
\end{acknowledgement}

\end{document}